\documentclass[11pt]{article}

\usepackage{graphicx}
\usepackage{latexsym}
\usepackage{fancyhdr}
\usepackage{subfigure}
\usepackage{mdwlist}

\makeatletter

\def\noflash#1{\setbox0=\hbox{#1}\hbox to 1\wd0{\hfill}}

\newcommand{\comment}[1]{}

\newcommand{\nocomment}[1]{}

\newcommand{\Iitemize}{\begin{itemize}
	{\setlength{\itemsep}{-6pt}}
       }

\newcommand{\ls}[1]
   {\dimen0=\fontdimen6\the\font 
    \lineskip=#1\dimen0
    \advance\lineskip.5\fontdimen5\the\font
    \advance\lineskip-\dimen0
    \lineskiplimit=.9\lineskip
    \baselineskip=\lineskip
    \advance\baselineskip\dimen0
    \normallineskip\lineskip
    \normallineskiplimit\lineskiplimit
    \normalbaselineskip\baselineskip
    \ignorespaces
   }

\def\ifundefined#1{\expandafter\ifx\csname#1\endcsname\relax}
\ifundefined{newblock}{
 }\fi

\newcommand{\eqref}[1]{Equation~\ref{#1}}

\begin{document}

\title{Short Note on Complexity of Multi-Value Byzantine Agreement \footnote{\normalsize This research is supported
in part by Army Research Office grant
W-911-NF-0710287. Any opinions, findings, and conclusions or recommendations expressed here are those of the authors and do not
necessarily reflect the views of the funding agencies or the U.S. government.}}

\date{\today}
\author{Guanfeng Liang and Nitin Vaidya\\ \normalsize Department of Electrical and Computer Engineering, and\\ \normalsize Coordinated Science Laboratory\\ \normalsize University of Illinois at Urbana-Champaign\\ \normalsize gliang2@illinois.edu, nhv@illinois.edu\\~\\Technical Report}

%


\maketitle


\thispagestyle{empty}

\newpage

\setcounter{page}{1}

\section{Introduction}

Inspired by \cite{multi-valued_BA_PODC06}, and the deterministic multi-valued Byzantine agreement algorithm in our recent technical report
 \cite{techreport_BA_complexity}, we derive a randomized algorithm that achieves multi-valued
Byzantine agreement with high probability, and achieves optimal
complexity. The discussion in this note is not self-contained, and relies heavily on the material in \cite{techreport_BA_complexity} -- please refer to
\cite{techreport_BA_complexity} for the necessary background.

Consider a synchronous fully connected network with $n$ nodes, namely $0,1,\dots, n-1$. Let node 0 be the source; the other $n-1$ nodes
are called {\em peers}. At most $t<n/3$ nodes can be faulty. The goal here is for the $n-1$ peers
to agree on the values sent by the source (similar to the Byzantine Generals problem in the work of Pease, Shostak and Lamport).
This is also known as the ``broadcast'' problem.
Our algorithm achieves agreement on a long message of $l$ bits with high probability. Similar to the algorithm in \cite{techreport_BA_complexity}, the proposed randomized Byzantine agreement algorithm  progresses in generations. In each generation, $D$ bits are being agreed upon, with the total number of generations being $l/D$.
For convenience, we assume $l$ to be an integral multiple of $D$.

\section{Operations when no failure detected}\label{sec:initial}
\paragraph{Step 1:} The source node 0 sends the $D$ bits to each of the peers. The peers do not transmit. 


\paragraph{Step 2:} Every node $i$ (including the source) computes $h_{i,j}=(K_{i,j},H(m_i,K_{i,j}))$ for every $j\neq i$, where (i) $K_{i,j}$ is a
randomly selected key of $k$ bits, (ii) for $i\neq 0$, $m_i$ represents the $D$ bits of the current generation
received by node $i$, and for $i=0$, $m_i$ represents the $D$ bits of data that the source sends in step 1, and (iii)
$H(m,K)$ is the almost-universal hash function
of $D/k$ bits introduced in \cite{Hirt07EfficientBA}. For convenience, we assume $D$ to be an integral multiple of $k$.
Then node $i$ sends $h_{i,j}$ to node $j$.

\paragraph{Step 3:} Every fault-free node $i$, on receipt of $h_{j,i}=(K_{j,i},h^*)$ from node $j$, computes hash of $m_i$ using key $K_{j,i}$
and compares it with $h^*$ in $h_{j,i}$. If any of these comparisons results in a mismatch, then node $i$ has detected inconsistency (or
misbehavior by a faulty node). After receiving the hash values from all the peers, each node $i$ broadcasts a 1-bit notification, indicating whether all hash values are consistent with $m_i$ or not, using a traditional Byzantine agreement algorithm -- similar approach
is used for  the deterministic agreement algorithm in \cite{techreport_BA_complexity}. If no notification indicating inconsistency is received, then every node $i$ decides on $m_i$, and the current generation completes. If any node indicates inconsistency detected, then the extended step (described below) is added. 

According to \cite{Hirt07EfficientBA}, the probability for any distinct $D$-bit messages $m_i$ and $m_j$ to produce the same hash value $H(m_i,K) =H(m_j,K)$ for a random key $K$ of $k$ bits is upper bounded by $2^{-k}D/k$. For the fault-free nodes to decide on different values, at least one comparison of hash values must erroneously result in a match. Thus, the probability that the fault-free nodes will decide on different values is upper bounded by $2^{-k}D/k$.  (A better bound on this probability can potentially be derived, but this bound suffices our purpose here.) Let us assume that
$2^{-k}D/k<1$, by proper choice of $k$.

\paragraph{Extended Step:} In the extended step, every node broadcasts all the packets it has received or has sent in steps 1 and 2, using a traditional Byzantine agreement algorithm. Using these broadcast information, identical ``diagnosis graphs'' are formed at all fault-free nodes. The formation and use of the diagnosis graph here is the same as the algorithm in \cite{techreport_BA_complexity}. The reader is referred
to \cite{techreport_BA_complexity} for the details of the diagnosis
graph.

\section{Operations after failure detected}\label{sec:detected}
After a failure is detected, and the extended step is finished, a new generation of $D$ bits of new data begins. Let us say that nodes $i$ and $j$ accuse(trust) each other if edge $ij$ is marked $f$($g$) in the diagnosis graph (see \cite{techreport_BA_complexity}).
Since a fault-free node never accuses another fault-free node, a fault-free node can be accused by at most $t$ other nodes.
If a node is accused by more than $t$ other nodes, this node is identified as faulty.
If the source node 0 is identified as faulty, then the fault-free peers can terminate the algorithm and all agree on some default value.
If a peer is identified as faulty, it is {\em isolated} (or removed) from the network, and
the algorithm below is executed only by the remaining nodes. Now consider the case when the source node is accused by no more than $t$ peers. 

\paragraph{Step 1:} Find a spanning tree routed at the source such that (i) the tree covers all nodes that have not been isolated, and (ii) it consists only $g$ edges in the diagnosis graph. Then the $D$ bits of data of the current generation is routed through this spanning tree.

Since the source node is not identified as faulty, it is connected to at least $n-t-1>1$ peers, each with a $g$-edge directly. Then, to show that the required spanning tree always exists, it only left to show that if the source and all $f$-edges are removed, the remaining nodes that have not been isolated are all connected to each other. Consider any pair of peers $i$ and $j$ that are not identified as faulty. Consider two cases:
\begin{itemize}
\item Nodes $i$ and $j$ trust each other: Then edge $ij$ between nodes $i$ and $j$ in the diagnosis graph must be a $g$-edge.
\item Nodes $i$ and $j$ accuse each other: Since any node that is not yet isolated can accuse at most $t$ nodes, 
nodes $i$ and $j$ each may accuse at most $t-1$ of the other $n-3$ peers.
Thus, among the other $n-3$ peers, node $i$ trusts at least $(n-3)-(t-1)=n-t-2 \ge (3t+1)-t-2 \ge 2t-1 \ge t$ peers (we assume $t\geq 1$).
%
%
Since node $j$ may accuse at most $t-1$ of the $t$ other peers that node $i$ trusts,
it follows that there exists at least one other peer that $i$ and $j$ both trust.
Thus, nodes $i$ and $j$ are connected in the diagnosis graph with a 2-hop path consisting
of g-edges.
\end{itemize}

The rest of the algorithm is the same as the case when no failure is yet detected.
As a clarification, note that in step 3, only nodes that are not isolated already
may send or receive messages.

\section{Security and Complexity Analysis}\label{sec:analysis}
\subsection{Security of the Algorithm}\label{subsec:security}

The security of the algorithm relies on the fact that the keys are not
sent in step 3 until step 2 is complete.
As seen in Section \ref{sec:initial}, the probability of the misbehavior by the faulty nodes being undetected is upper bounded by $2^{-k}D/k$.
 In other words, the misbehavior in a particular generation will be detected with probability at least  $\rho = 1-2^{-k}D/k$. 
Then the probability that the misbehavior is always detected given that the faulty nodes misbehave in $x$ generations is 
lower bounded by $\rho^x$, which is a decreasing function in $x$. Notice that if first $t(t+1)$ instances of misbehavior
is detected, then all faulty nodes will be identified and isolated (by an ``instance'' we mean a generation in which 
at least one faulty node misbehaves by sending inconsistent data). Thus, the probability that the misbehavior is always
detected, i.e.,  the probability of achieving agreement correctly on all $l$ bits, is 
\begin{eqnarray}
P_{correct} \geq \rho^{t(t+1)}&=&(1-2^{-k}D/k)^{t(t+1)} \\
&=& 1 - O(t(t+1)2^{-k}D/k)) \\
&=& 1 - O\left(n^2\,2^{-k}D/k\right) \label{eq:security}
\end{eqnarray}

\subsection{Complexity of the Algorithm}\label{subsec:complexity}
\paragraph{Data transmissions:} Every peer receives $D$ bits through steps 1 and 2. So $(n-1)D$ bits are transmitted in each generation, which leads to $(n-1)l$ bits for data transmissions throughout the whole algorithm.

\paragraph{Hash keys:} Every node that is not identified as faulty sends $k+D/k$ bits ($k$ bit key, and $D/k$ bit hash value) to every other node. So at most $n(n-1)(k+D/k)$ bits are transmitted in each generation, which leads to at most $n(n-1)(k+D/k)l/D$ bits throughout the whole algorithm.

\paragraph{Broadcasts in step 4:} In step 4, every node broadcasts a 1 bit of notification. Let us denote $B$ as the communication complexity of broadcasting 1 bit. Then the total cost for the broadcast in step 4 is $nB$ bits, which lead to $nBl/D$ bits over the whole algorithm.

\paragraph{Broadcasts in extended step:} In the extended step, every node broadcasts $D$ bits. Thus $nDB$ bits are transmitted in each extended step. Since there will be at most $t(t+1)$ extended steps, the total cost of broadcasts in extended steps is at most $nDBt(t+1)$ bits throughout the whole algorithm.

Now we have a upper bound on $C(l)$, the total number of bits being transmitted to achieve agreement on $l$ bits, as:
\begin{eqnarray}
(n-1)l + n(n-1)(k+D/k)l/D + nBl/D + nDBt(t+1)\\
 = (n-1)l + O(n^2kl/D + n^2l/k+ nBl/D + n^3BD).
\end{eqnarray}
Notice that broadcast algorithm of complexity $\Theta(n^2)$ are known \cite{bit_optimal_89,opt_bit_Welch92}, so we assume $B=\Theta(n^2)$. Then we have 
\begin{equation}
C(l) = (n-1)l +  O(n^2l/k + (n^2k + n^3)l/D + n^5D).
\end{equation}
Then the per-bit communication complexity is
\begin{equation}
\alpha = C(l)/l = n-1 + O(n^2/k + (n^2k + n^3)/D + n^5D/l). \label{eq:complexity}
\end{equation}

\subsection{Achieving high probability of agreement with low per-bit complexity}
Now let us consider Equations \ref{eq:security} and \ref{eq:complexity} together. If we choose $k$ and $D$ such that the following conditions are all satisfied
\begin{itemize}
\item $k$ and $D$ are both unbounded increasing functions of $l$;
\item $D = o(2^k k)$ and $D = o(l)$;
\item $k = o(D)$,
\end{itemize}
then $P_{correct}\rightarrow 1$ and $\alpha \rightarrow n-1$ as $l$ gets large. For example, we can choose $k=\log l$ and $D = l^{1-\beta}$ for some positive constant $0<\beta<1$, then 
\begin{equation}
P_{correct} = 1 - O(n^2l^{-\beta}/\log l)
\end{equation}
\begin{equation}
\alpha= n-1 +   O(n^2/\log l+(n^2\log l  + n^3 )l^{-(1-\beta)} + n^5 l^{-\beta}).
\end{equation}

From \cite{multi-valued_BA_PODC06,techreport_BA_complexity}, we know that the per-bit complexity $\alpha$ is lower bounded by $n-1$. Thus, as $l$ approaches $\infty$, the proposed algorithm achieves
agreement of $l$ bits with probability approaching 1, with per-bit complexity approaching the lower bound of $(n-1)$.

\bibliographystyle{abbrv}
\bibliography{PaperList}

\begin{thebibliography}{1}

\bibitem{Hirt07EfficientBA}
Z.~Beerliova-Trubiniova, M.~Hirt, and M.~Riser.
\newblock {\em Efficient Byzantine Agreement with Faulty Minority}.
\newblock Springer-Verlag, 2007.

\bibitem{bit_optimal_89}
P.~Berman, J.~A. Garay, and K.~J. Perry.
\newblock Bit optimal distributed consensus.
\newblock {\em Computer science: research and applications}, 1992.

\bibitem{opt_bit_Welch92}
B.~A. Coan and J.~L. Welch.
\newblock Modular construction of a byzantine agreement protocol with optimal
  message bit complexity.
\newblock {\em Inf. Comput.}, 97(1):61--85, 1992.

\bibitem{multi-valued_BA_PODC06}
M.~Fitzi and M.~Hirt.
\newblock Optimally efficient multi-valued byzantine agreement.
\newblock In {\em PODC '06}, 2006.

\bibitem{techreport_BA_complexity}
G.~Liang and N.~Vaidya.
\newblock Complexity of multi-valued byzantine agreement.
\newblock {\em Technical Report, CSL, UIUC}, June 2010.

\end{thebibliography}

\end{document}